\documentstyle[epsfig,prl,aps,twocolumn]{revtex}
\begin{document}

\wideabs{
\title{New Phases of Solid Nitrogen}

\author{Eugene Gregoryanz$^1$\cite{byline}, Alexander Goncharov$^1$ 
\cite{byline}, Russell J. Hemley$^1$, Ho-kwang Mao$^1$, Maddury Somayazulu$^2$ 
and Guoyin Shen$^3$}

\address{$^1$Geophysical Laboratory and Center for High
Pressure Research, Carnegie Institution of Washington, \\
5251 Broad Branch Road NW, Washington D.C. 20015 U.S.A \\
$^2$ HPCAT, Carnegie Institution of Washington,  ANL,
9700 South Cass Avenue, Argonne, IL   60439 \\
$^3$Consortium for Advanced Radiation Sources, 
University of Chicago, 9700 South Cass Avenue, Argonne, IL 60439 } 
\maketitle 

\begin{abstract}
We report the discovery of a new class of molecular phases of solid
nitrogen at high pressures and temperatures by Raman and infrared
spectroscopy and powder x-ray diffraction. Unlike the molecular phases
consisting of disk- and sphere-like molecular disorder ($\delta$-, $\epsilon$-
and $\zeta$-N$_2$) and reportedly stable over a wide P-T range, one of the new
phases ($\iota$) is diatomic with disk-like molecules. A second new phase
(the higher pressure $\theta$-phase) is characterized by strong
intermolecular interactions and infrared vibron absorption. Both phases
exhibit wide P-T ranges of stability and metastability.
\end{abstract}   
}

The evolution of
molecular solids under pressure constitutes an important problem
of modern physics \cite{hemley}. Under compression, delocalization
of electronic shells and eventual molecular dissociation
with formation of a framework or closed packed structures is
expected. This pathway is not necessarily a straightforward
process, because of large barriers of transformation between
states with different types of bonding and molecular structures
with various types of orientational order, including possible
dimers or associated and charge transfer compounds. Nitrogen
is an archetypal homopolar diatomic molecule with a very
strong intramolecular bonding. The phase diagram of nitrogen
is complex at moderate pressures and temperatures and has
been little studied over a wider range  until recently \cite{bini}. A
theoretically proposed dissociation of nitrogen molecules
under pressure \cite{mcmahan} recently was confirmed experimentally
\cite{goncharov,eremets}. Our previous study \cite{gregoryanz} 
suggests that the nonmolecular
material obtained on cold compression is amorphous, formed as
a consequence of the large barrier of transformation to the
crystalline phase.  Here, we report the existence of two new
molecular phases, called  $\iota$ and $\theta$, which have exceptionally
large region of stability and metastability extending  through
the P-T region  where $\epsilon$ and $\zeta$ have been thought to be the only
stable phases of nitrogen. This observation establishes a new
class of dense solid molecular N$_2$ phases.

The phase diagram of nitrogen is shown in Fig. 1. The
melting curve  was measured up to 900 K \cite{young}; from liquid phase,
which exists to 2 GPa at room temperature, nitrogen solidifies
in a disordered (plastic) $\beta$ phase, which is a common feature of
diatomic molecular crystals \cite{freiman}. Low pressure and temperature
$\alpha$ and $\gamma$ phases represent two ways of packing quadrupoles
\cite{freiman}. At higher pressures, another class of structures with
non quadrupolar-type ordering was discovered ($\delta$, $\delta_{loc}$,
$\epsilon$, $\zeta$) \cite{lesar,schiferl,scheerboom,cromer,mills}. 
This happens because the relative contribution of
the quadrupolar interactions gradually decrease, and other
interactions (e.g., short-range repulsive) come to play.
The $\delta$ phase  is disordered  with sphere- and disk-like types
of molecules orientationally distributed between corners
and faces of a fcc unit cell (space group Pm3n) \cite{cromer}. With
decreasing temperature and/or increasing pressure molecules
partially order in $\delta_{loc}$-N$_2$  phase and then completely in 
$\epsilon$-N$_2$. The latter has a distorted cubic (rhombohedral space group
R3c) structure as determined from in situ x-ray measurements
\cite{mills}. Vibrational spectroscopy shows that this phase preserves
the difference between sphere- and disk-like molecules. Further
increase  in pressure  leads to several sequential changes
in vibrational spectra, namely increase of a number of the
lattice modes and branching of vibron mode corresponding to
disk-like molecules \cite{reichlin,olijnyk} . With a lack of decisive x-ray
data \cite{jephcoat}, these transformations have been  interpreted as due
to a further lowering of symmetry \cite{schiferl}. The available
data indicate that only $\zeta$ phase  can  be considered as well
established molecular phase at higher pressures \cite{olijnyk}. The
transformation to this phase  is quite pronounced at 21-25 GPa
at low temperatures \cite{schiferl}, but at room temperature the changes in
vibrational spectra are relatively subtle \cite{reichlin,olijnyk}. 
Theoretical calculations confirm the stability of $\epsilon$ 
phase as the ordered phase from 2 to 41.5 GPa  \cite{etters}, 
although  simulations \cite{belak} favor tetragonal multimolecular 
structures, which disagree with the known experimental data at 
the same pressure range.

The P-T region, of the phases derived from $\delta$-N$_2$ 
($\delta_{loc}$, $\epsilon$, $\zeta$)
are observed is quite wide (Fig. 1), apparently  extending  to
the transition to nonmolecular phase $\eta$ \cite{goncharov,gregoryanz}. 
For comparison, $\gamma$-O$_2$,  a structural analog of $\delta$-N$_2$ 
(see also $\beta$-F$_2$ and CO \cite{cromer}) is stable in 
a relatively narrow PT range, beyond which
structures with collinear molecules become energetically
favorable (O$_2$,F$_2$) \cite{cromer} or chemical dissociation takes place
CO). Although the above molecular crystals are different from
the point of view of anisotropic intermolecular forces \cite{cromer},
one would expect that at high density the  most effective
packing of dumbbell molecules remains the  dominant term in
determining phase  stability prior to dissociation.  

Prior to the transition to nondiatomic phase, the Raman
vibron shows a substantial softening \cite{reichlin,olijnyk,jephcoat}
 which can arise from weakening of the intramolecular bond but can also arise
from increased vibrational splitting. Information about the
vibron frequencies and their pressure dependence in different
ordered molecular phases is important for understanding of
the nature of this ordering. For example, an abrupt drop in
the vibrational frequency at a structural transition can be
due to delocalization of the intramolecular bonding charge
giving possible dimer formation \cite{gorelli} and/or charge transfer
phases \cite{kohanoff}. At the transition to the nonmolecular phase, the
lattice excitations and vibrons are replaced by new excitations,
consistent with nondiatomic  structure \cite{goncharov,gregoryanz}. 
Detailed study
of vibrational properties and structure of semiconducting
absorption edge suggests disorder, at least for certain P-T
paths followed to create the material \cite{gregoryanz}.  In order to
establish the relation between molecular and non-molecular
phases we examined nitrogen from 15 to 1000 K and up to 150
GPa using different experimental techniques including in situ
high-temperature visible and Raman spectroscopy as well IR
\cite{goncharov1}synchrotron spectroscopy and x-ray synchrotron diffraction
used on samples quenched to the room temperature (RT).

When compressed at RT nitrogen transforms from the $\epsilon$
 to $\zeta$ phase
around 60 GPa (see Fig. 1) \cite{goncharov2}.  When heated, sample first
back transforms from $\zeta$ to $\epsilon$ phase along the boundary, which we
find to be on the extension of the line established in Ref. \cite{bini}
at lower temperatures. When temperature reaches $\sim$625 K phase
transition to $\theta$ nitrogen takes place. The transition can be
observed visually (see inset to Fig. 1) since  - N$_2$ normally
shows a substantial  grain boundaries, while after transition
to the $\theta$  phase, the sample looks uniform and translucent. In
most cases the transition happens instantaneously and completely
(determined by Raman spectroscopy) within seconds. If 
$\epsilon$- N$_2$ is
heated at even lower pressures (e.g., 60-70 GPa),  it transforms
above 750 K to $\iota$- N$_2$. It is also possible to access $\iota$ phase from
$\theta$. In one of our experiments we observed the transformation from
the $\theta$  to $\iota$  phase on pressure release at $\sim$850 K at 69 GPa.

Fig. 2 shows the Raman and IR absorption spectra of the
$\theta$ and $\iota$ phases quenched to room temperature. The spectrum of
the $\zeta$ phase (initial material) obtained by "cold" compression
is shown for comparison. The Raman and IR spectra of both the
$\theta$ and $\iota$ phases exhibit vibron modes, although their number and
frequencies differ from those  of all other known molecular
structures (see below). The infrared vibron mode of the q phase
has much larger oscillator strength compared to other N$_2$ phases
(cf. H$_2$ in phase III \cite{hanfland} and $\epsilon$-O$_2$ 
\cite{gorelli}). The lattice modes of
$\theta$ nitrogen are very sharp compared to either $\iota$ or 
$\zeta$ ($\epsilon$). This
is a clear indication that molecular ordering in $\theta$ phase is
essentially complete, whereas other molecular phases still
possess substantial amount of static or dynamic orientational
disorder. Comparison of Raman and infrared vibron modes shows
a general softening of the vibron bands of $\theta$ nitrogen compared
to the other modifications ($\iota, \zeta, \epsilon$). This fact and also a
presence of relatively strong infrared vibron band  indicate
a charge transfer from intra- to intermolecular bonds and
strengthen our arguments about ordered nature of the phase
\cite{gorelli,kohanoff}. Raman and infrared spectra show several cases of
frequency coincidence  of Raman and infrared vibron and lattice
modes, which excludes an inversion center for both structures.

The pressure dependence of the Raman-active vibron modes
(Fig. 3) was studied on unloading at 300 K in both phases
(see discussion below). $\iota$-N$_2$ exhibits a typical behavior
for molecular crystals: branching of vibrational modes and
increasing of separation between them with pressure due to
increasing of intermolecular interactions. All the vibrational
modes originate from the same center, which is close to the
frequency  of the $\nu_2$ disk-like molecules in  $\epsilon$
-N$_2$. Thus,
the structure of the $\iota$ phase is characterized by presence of
just one type of site symmetry occupied by the molecules,
and a large number of vibrational modes arise from a large
unit cell (a minimum 8 molecules per cell). For the $\theta$ phase,
two different site symmetries appear to be occupied. The
higher frequency  $\nu_{1\theta}$ gives rise to three Raman bands and one
IR, while the lower frequency of $\nu_{2theta}$ - correlates  with only
one Raman band.. Detailed analysis of vibratioanal spectra
of both $\theta$ and $\iota$ phases will be published elsewhere.

Synchrotron  x-ray diffraction data (Fig. 4) confirm the
existence of two new structures. First, the data  show a good
agreement with previously reported results  for the $\epsilon$-phase
\cite{olijnyk1,hanfland1}. Only a few reflections could be observed above 50
GPa because of a strong sample texture. No major changes
in the x-ray diffraction patterns were observed at 60 GPa
and room temperature, corresponding to the $\epsilon-\zeta$ transition
(see also Ref. \cite{jephcoat}). This observation is consistent with
vibrational spectroscopy, which shows  only  moderate changes
identified  as a further distortion of the cubic unit cell
of the $\delta$ phase \cite{reichlin,goncharov2}. In contrast, 
the x-ray diffraction
patterns of the samples after $\zeta-\theta$ and $\zeta-\iota$
 transformations
differ substantially from those of the $\epsilon$ and  $\zeta$  phases,
and from each other (Fig. 4). Indexing of the peaks of $\theta$ 
nitrogen shows that its unit cell is  orthorhombic,  for
example with the lattice parameters a=6.797(4), b=7.756(5)
and  c=3.761(1). The systematic absences, lack of inversion
center and presence of high-symmetry sites (see above) are
consistent with the space groups Pma2, Pmn21, Pmc21, Pnc2,
P21212. The a/c ratio is close to $\sqrt{3}$, which clearly suggests
that the lattice is  derived from a  hexagonal structure
(cf. hydrogen in phase III). Extrapolation of the equation
of state of $\epsilon$- N$_2$ measured to 40 GPa [24] shows that the
molecular volume for this phase is about 14 $\dot{A}^{3}$/molecule at
95 GPa, which gives an upper bound assuming a pressure-induced
(density driven) transition. Comparison with the experimentally
determined unit cell volume (198 $\dot{A}^{3}$) suggests 16 molecules in
the unit cell, giving  12.4 $\dot{A}^3$ per molecule in the 
$\theta$-phase
and 11\% volume collapse at the $\epsilon-\theta$ transition. The number of
molecules is in agreement with	vibrational spectroscopy data,
although it is possible to describe the vibrational spectra
with smaller number (up to 8).   

 In order to better
understand  the provenance of new phases in the phase diagram
and their relation with other phases of nitrogen (Fig. 1),
we pursued extensive observations in different parts of the
phase diagram. The new phases have a wide range of stability
or metastability. As it was noted above, both phases could be
quenched to room temperature. On subsequent heating, the $theta$ phase
remained stable when heated above 1000 K between 70-135 GPa,
but it transforms to $\iota$-N$_2$ when releasing pressure at 68 GPa at
850 K (see above). In view of the relatively high temperature
of this transformation and its absence at room temperature,
this observation implies that  transformation point is close
to the thermodynamical	$\theta-\iota$ transition (see Fig. 1). At room
temperature,  $\theta$ nitrogen remains metastable as low as 30 GPa on
unloading. Similarly, the $\iota$-N$_2$ remains metastable to 23 GPa; at
these pressures both phases transform to $\epsilon$-N$_2$ on unloading. 
$\iota$- N$_2$ was found to be stable at low temperatures (down to 10
K) at pressures as low as 30 GPa. It is interesting to note
that amorphous $\eta$ nitrogen can be accessed only from $\zeta$-N$_2$ (see
Fig. 1). The apparent kinetic boundary, that separates these
phases can be treated as a line of instability of $\zeta$-N$_2$. On
the other side, the $\iota$ and $\theta$ phases can be reached from 
$\epsilon$ (and
maybe $\delta$) phase only. We observed that on further increasing
of pressure and temperature $\theta$ phase does not transform to the
nonmolecular phase $\eta$  (to at least 135 GPa and 1050 K). We
suggest that, it might instead transform to a (perhaps
different) nonmolecular crystalline phase on compression.
By this argument, this phase cannot be reached by "cold"
compression because of the kinetic barrier separating it from
$\zeta$-N$_2$. Also, one can speculate that heating of the amorphous
phase can help to overcome the energy barrier and induce
its crystallization.   

The data presented allow us to
speculate about the nature of $\iota$ and $\theta$ nitrogen and their large
metastability on unloading. The $\iota$-phase consists of disk-like
molecules, presumably packed more efficiently compared to the
mixed disk- and sphere-like $\delta$-family structures. The 
$\theta$-phase
is more complex; it is very tempting to interpret this phase
as due to formation of polyatomic molecules (e.g. N$_2$-N$_5$),
as has been suggested by theoretical calculations \cite{bartlett}.
The striking vibrational properties of $\theta$  nitrogen strongly
suggest  some kind of associated phase , perhaps with some
analogy to  H$_2$-III \cite{hemley1}, $\epsilon$-O$_2$ \cite{gorelli}
 or CO$_2$-II \cite{iota}. The
phase diagram of nitrogen appears to be very complex. The
presence of substantial barriers of transformations makes
the transformations very sluggish. Thus, determination of the
thermodynamical boundaries between phases requires additional
studies.  The data obtained in this study clearly show that
the new phases are thermodynamically stable high-pressure
phases. This conclusion is based on the fact that the same phase
is formed irrespective of the thermodynamic path. Moreover,
the  $\zeta$-N$_2$ seems to be metastable, since it can be obtained
only as a result of "cold" compression of the $\epsilon$-N$_2$ (see also
Refs. \cite{gregoryanz,goncharov2}.   

The authors are grateful to Y. Fei
for the help with high-temperature experiment, P. Dera for
help with analyzing x-ray diffraction patters and Z. Liu for
help with IR experiments. 
Synchrotron x-ray diffraction was performed at the GSECARS sector at the
Advanced Photon Source.  GSECARS sector is supported by NSF (Earth Science
Instrumentation and Facilities Program), DOE (Geoscience Program) and W. M.
Keck Foundation.

%%%%%%%%%%%%%%%%%%%%%%%%%%%%references%%%%%%%%%%%%%%%%%%%%

%%%%%%%%%%%figures%%%%%%%%%%%%%%%
\newpage

\begin{figure}
\centerline{\epsfig{file=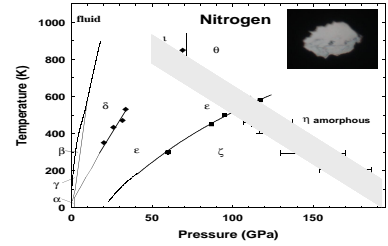,width=6.cm,height=6.cm}}
\caption{ Phase diagram of nitrogen. Filled
symbols and thick solid lines are the data from this work and
Refs. 6,22. Dashed-dotted line is the proposed extension of
the $\zeta-\eta$ transition boundary (see text). Open circles and short
dashed line are from visual observations of Ref. 5. Phase
boundaries at low pressures are from Refs. 2,7.  The phase
boundaries for $\alpha, \gamma, \delta and \delta_{loc}$ 
phases are not shown. The inset shows the photo of 
theta phase at 95 GPa and 300 K.}  
\label{fig1}
\end{figure}

\newpage

\begin{figure}
\centerline{\epsfig{file=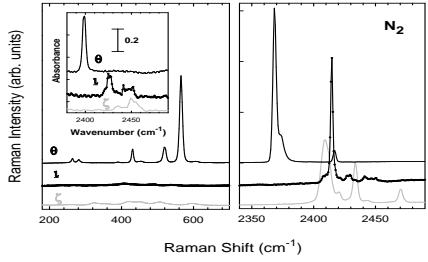,width=8.cm,height=8.cm}}
\caption{Representative Raman and IR spectra (inset)
of $\theta$ (solid lines) and $\iota$ (lines with dots) phases measured at
95 and 70 GPa and 297 K upon quenching from high temperature
(see text). The spectra of $\zeta$ phase (gray lines) are shown for
comparison.}
\label{fig2}
\end{figure}  

\begin{figure}
\centerline{\epsfig{file=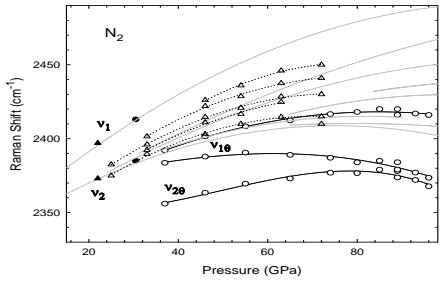,width=8.cm,height=8.cm}}
\caption{ Raman frequencies of vibron
modes as a function of pressure for $\theta$ (open circles and solid
lines) and $\iota$ (open triangles and dashed line) phases measured at
the pressure release. Filled circles and triangles correspond
to the vibron frequencies after transformation to the $\epsilon$ phase
from $\theta$ and $\iota$ phases, respectively. Gray lines are 
data for $\epsilon(\zeta$) phases from Ref. 15.}
\label{fig3}
\end{figure}

\begin{figure}
\centerline{\epsfig{file=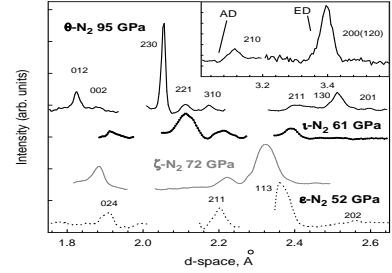,width=8.cm,height=8.cm}}
\caption{X-ray diffraction patterns of $\theta$ (solid lines), 
$\iota$ (lines with dots), $\zeta$ phase (gray
lines) and $\epsilon$ (dashed lines) measured at different pressures
and 297 K. AD and ED stand for angle- and energy-dispersive
techniques, respectively. The diffraction lines corresponding
to rhenium were subtracted. The corresponding d-space ranges
are omitted. Indexing of $\epsilon$-N$_2$ pattern is according to
Ref. 24.}  
\label{fig4}
\end{figure}

\end{document}